\documentclass[a4paper,fleqn,usenatbib]{mnras}
\usepackage{txfonts}

\usepackage[T1]{fontenc}
\usepackage{ae,aecompl}

\usepackage{graphicx}	
\usepackage{amssymb}	

\title[The subgiant branch of NGC 411]{The tight subgiant branch of the intermediate-age star cluster NGC 411 implies a single-aged stellar population}
\author[C. Li et al.]{C. Li$^{1,2,3,4}$\thanks{email: licy@pmo.ac.cn}, R. de Grijs$^{2,5}$, N. Bastian$^{6}$, L. Deng$^{3}$, F. Niederhofer$^{7,8,9}$ and\newauthor C. Zhang$^{10}$
\\
$^{1}$ Purple Mountain Observatory, Chinese Academy of
  Sciences, Nanjing 210008, China\\
$^{2}$ Kavli Institute for Astronomy \& Astrophysics and
  Department of Astronomy, Peking University, \\Yi He Yuan Lu 5, Hai
  Dian District, Beijing 100871, China\\
$^{3}$ Key Laboratory for Optical Astronomy, National
  Astronomical Observatories, Chinese Academy of Sciences, \\ 20A Datun
  Road, Chaoyang District, Beijing 100012, China\\
$^{4}$ Department of Physics and Astronomy, Macquarie
  University, Sydney, NSW 2109, Australia\\
$^{5}$ International Space Science Institute--Beijing, 1
  Nanertiao, Zhongguancun, Hai Dian District, Beijing 100190, China\\
$^{6}$ Astrophysics Research Institute, Liverpool John
  Moores University, 146 Brownlow Hill, Liverpool L3 5RF, UK\\
$^{7}$ Excellence Cluster `Origin and Structure of the
  Universe', Boltzmannstra{\ss}e 2, D-85748 Garching bei M\"unchen, Germany\\
$^{8}$ Universit\"ats-Sternwarte M\"unchen,
  Scheinerstra{\ss}e 1, D-81679 M\"unchen, Germany\\
$^{9}$ Space Telescope Science Institute, 3700 San Martin Drive, 
Baltimore, MD 21218, USA\\
$^{10}$ Shanghai Astronomical Observatory, Chinese Academy of
  Sciences, Shanghai 200030, China\\}

\date{Accepted XXX. Received YYY; in original form ZZZ}

\pubyear{2016}

\begin{document}
\label{firstpage}
\pagerange{\pageref{firstpage}--\pageref{lastpage}}
\maketitle

\begin{abstract}
The presence of extended main-sequence turn-off (eMSTO) regions in
intermediate-age star clusters in the Large and Small Magellanic
Clouds is often interpreted as resulting from extended star-formation
histories (SFHs), lasting $\geq$ 300 Myr. This strongly conflicts with
the traditional view of the dominant star-formation mode in stellar
clusters, which are thought of as single-aged stellar
populations. Here we present a test of this interpretation by
exploring the morphology of the subgiant branch (SGB) of NGC 411,
which hosts possibly the most extended eMSTO among all known
intermediate-age star clusters. We show that the width of the NGC 411
SGB favours the single-aged stellar population interpretation and
rules out an extended SFH. In addition, when considering the
red clump (RC) morphology and adopting the unproven
premise that the widths of all features in the colour--magnitude
diagram are determined by an underlying range in ages, we find that
the SFH implied is still very close to that resulting from a
single-aged stellar population, with a minor fraction of stars
scattering to younger ages compared with the bulk of the
population. The SFHs derived from the SGB and RC are both inconsistent
with the SFH derived from the eMSTO region. NGC 411 has a very low
escape velocity and it has unlikely undergone significant mass loss at
an early stage, thus indicating that it may lack the capacity to
capture most of its initial, expelled gas from stellar evolutionary
processes, a condition often required for extended SFHs to take root.

\end{abstract}


\section{Introduction}
\label{sec:intro}

Until about a decade ago, it was believed that -- except for the
oldest globular clusters -- star clusters contain stars described by
the theoretical concept of `simple' stellar populations (SSPs). They
were each thought to have formed in a particular starburst event, with
a maximum age spread of only 1--3 Myr \citep{Long14}. However, the
discovery of extended main-sequence turn-off (eMSTO) regions in
intermediate-age, 1--2 Gyr-old star clusters in the Large and Small
Magellanic Clouds (LMC, SMC) has radically changed our view and
understanding of the validity of the SSP scenario for stellar clusters
\citep[e.g.,][]{Mack07,Glat08,Mack08,Milo09,Kell12,Goud14}, because
the areas in colour--magnitude space covered by their eMSTO regions
are significantly more extended than would be expected for SSPs. A
direct interpretation of these eMSTOs in intermediate-age star
clusters, adopting the common yet unproven assumption that the eMSTO
width is entirely owing to a range in stellar ages, would imply that
these objects may have experienced extended star-formation histories
(SFHs) of {\it at least} 300 Myr. Since such features are found in
almost all intermediate-age star clusters \citep{Milo09}, this would
then strongly indicate that star formation can ubiquitously proceed
for several hundred million years within massive clusters. This has
led to extensive debates as to whether star clusters can harbour
extended SFHs.

\cite{Goud14} proposed a scenario that invokes two distinct rounds of
SFH in massive clusters to explain the observed eMSTO
regions. In their model, the first star-formation episode
  is a near-instantaneous burst, which is followed by a 10--100 Myr
  period with no star formation. In turn, this quiescent period is
  followed by a second star-formation episode that lasts a few hundred
  million years. \cite{Goud14} found that the width of the eMSTO of
intermediate-age star clusters is correlated with their central escape
velocity, $v_{\rm esc}$, with a threshold of 12--15 km s$^{-1}$ for
$v_{\rm esc}$. However, critiques of various aspects of this scenario
have been presented elsewhere
\citep{Bast15,DAnto15,Nied15b,Cabr15}. If \cite{Goud14} are on the
right track, the extended SFH scenario should be reflected by the
presence of age spreads in young massive clusters. Explorations of the
reality of such proposed dramatic age spreads in young massive
clusters have been undertaken by various teams based on either
analyses of cluster colour--magnitude diagrams \citep[CMDs;
  e.g.,][]{Lars11,Nied15}) or integrated spectroscopy
\citep{Bast13,Cabr14}. However, no significant age spreads have been
found to date.

At the same time, rapid stellar rotation has been identified as a
condition that may affect the area of the MSTO region as strongly as
stellar age spreads. This is because the centrifugal force somewhat
reduces stellar self-gravity, which leads to a reduction in the
stellar luminosity and surface temperature
\citep{Bast09,Lang12}. However, this model has been criticized by
\cite{Gira11}, who realized that the effect of a prolonged stellar
main-sequence (MS) lifetime owing to stellar rotational mixing
\citep[e.g.,][]{Meyn00} would cancel the broadening of MSTO regions
caused by this reduced self-gravity and, therefore, conspire to retain
a narrow MSTO region at a certain age. On the other hand,
\cite{Yang13} showed that if one adopts a reasonable significance for
the effects of this prolonged stellar lifetime, rapid stellar rotation
could still be a solution to the `multiple-age conundrum'; their
conclusion was recently supported by \cite{Bran15} and \cite{Nied15b}.

\cite{Li14} first addressed an new front in research by
exploring other parts of the CMD to constrain the internal age spreads
of intermediate-age star clusters. They found that the 1.7 Gyr-old
cluster NGC 1651, which hosts an eMSTO which would suggest the
presence of a 450 Myr age spread, possesses a very tight subgiant
branch (SGB) that favours a single-aged stellar population. They
affirmed that the cluster's tight SGB offers the strongest evidence
yet in support of the SSP scenario. Meanwhile, their idea was
independently confirmed by \cite{Bast15} for the intermediate-age
clusters NGC 1806 and NGC 1846. \cite{Nied15c} reached the same
conclusion based on a comparison of the extensts of the red clumps
(RCs) and the eMSTOs in 12 LMC clusters. However, based on more recent
calculations considering the effects of convective overshooting,
\cite{Goud15} showed that the SGB widths of NGC 1651, NGC 1806 and NGC
1846 might still be consistent with predictions involving significant
age spreads. \cite{Goud15} argued that the SGB locus in the CMD as
used by \cite{Li14} and \cite{Bast15} may not be the best choice for
further study; instead, they argued that one should consider the SGB 
morphology or `spread', which should be wide if age spreads are present 
and narrow if not. Indeed, there are compelling reasons to explore the 
structure of the SGB in intermediate-age star clusters to constrain their internal age
spreads, rather than solely concentrating on eMSTO regions. Finally,
\cite{Nied15c} showed that by taking into account the morphology of
the RC, the maximum viable age dispersion can be further
constrained. In most cases, the SFHs derived from RCs are smaller than
those apparently implied by the eMSTO regions.
  
In this paper, we focus on the SMC cluster NGC 411, whose MSTO region
is much more extended than those of any other intermediate-age star
clusters. Its eMSTO was first identified by \cite{Gira13}, who showed
that it is consistent with an age spread of $\sim$ 700 Myr, while
\cite{Goud14} even determined that the age spread of NGC 411 can reach
as much as 1 Gyr (their fig. 2). The mass of NGC 411 is only 32,000
M$_{\odot}$, making it one of the lowest-mass clusters hosting an
eMSTO, in apparent contradiction to \cite{Goud14}. In Section \ref{S2}
we introduce the details of our data reduction. In Section \ref{S3} we
present the main results. In Section \ref{S4}, we explore applicability of both 
the age-spread scenario and the rapid stellar rotation model to NGC 411. 
Our conclusions are presented in Section \ref{S5}.

\section{Data Reduction and Analysis}\label{S2}

Our data sets were obtained from {\sl Hubble Space Telescope} ({\sl
  HST}) programme GO-12257 (principal investigator: L. Girardi), using
the Wide Field Camera-3 (WFC3). The resulting data set is composed of
a combination of four science images in the F475W and F814W filters,
which roughly correspond to the Johnson--Cousins $B$ and $I$ bands,
respectively. The total exposure times for these two bands are 1520 s
and 1980 s, respectively.

The adopted photometric procedures are identical to those used by
\cite{Li15}, who based their results on two independent software
packages, DAOPHOT \citep{Davi94} and DOLPHOT
\citep{Dolp13}. The resulting stellar catalogues adopted here are
those from the DAOPHOT photometry. We confirmed that our results
are internally consistent by comparing them with the DOLPHOT
analysis.

In \cite{Li15} we determined that the cluster's number density centre
is $\alpha_{\rm J2000} = 01^{\rm h}07^{\rm m}56.22^{\rm s}$,
$\delta_{\rm J2000} = -71^{\circ}46'04.40''$. Its radial profile shows
that at a radius of 25 arcsec the stellar number density decreases to
about half the central density. To avoid background contamination as
much as possible, while keeping our analysis results statistically
robust, we select twice this value (50 arcsec) as the cluster's
typical region of interest.

At the distance of the SMC, we cannot use proper-motion selection to
decontaminate the background stars from the cluster sample.
Therefore, we adopted a statistical method to reduce background
contamination. We assume that the CMD of the field stars in a nearby
region should be similar to that in the cluster region. We then
generated numerous grids with sizes of 0.50$\times$0.25 mag$^2$ to
cover the entire CMD region. We randomly removed a number of stars
corresponding to that in the area-corrected field-region CMD from the
cluster CMD. Our method is similar to that employed by \cite{Hu10} and
\cite{Li13}. We confirmed that the adopted grid size will not affect
the observed features; viable cell sizes range from roughly
0.3$\times$0.15 mag$^2$ to 0.5$\times$0.25 mag$^2$. The observed
features will change significantly only if we adopt grid sizes that
are either too small (e.g., smaller than 0.1$\times$0.05 mag$^2$) or
too large (e.g., larger than 1.0$\times$0.50 mag$^2$). This is because
the relative number density of the NGC 411 SGB region is moderate; a
grid size which is too small would not contain sufficient numbers of
stars, while a much larger grid will smooth the observed details. The
details of our adopted field region are discussed in \citep[][see the
  right-hand panel of their Extended Data fig. 2]{Li15}. 

\begin{figure*}
\centering
\includegraphics[width= 2\columnwidth]{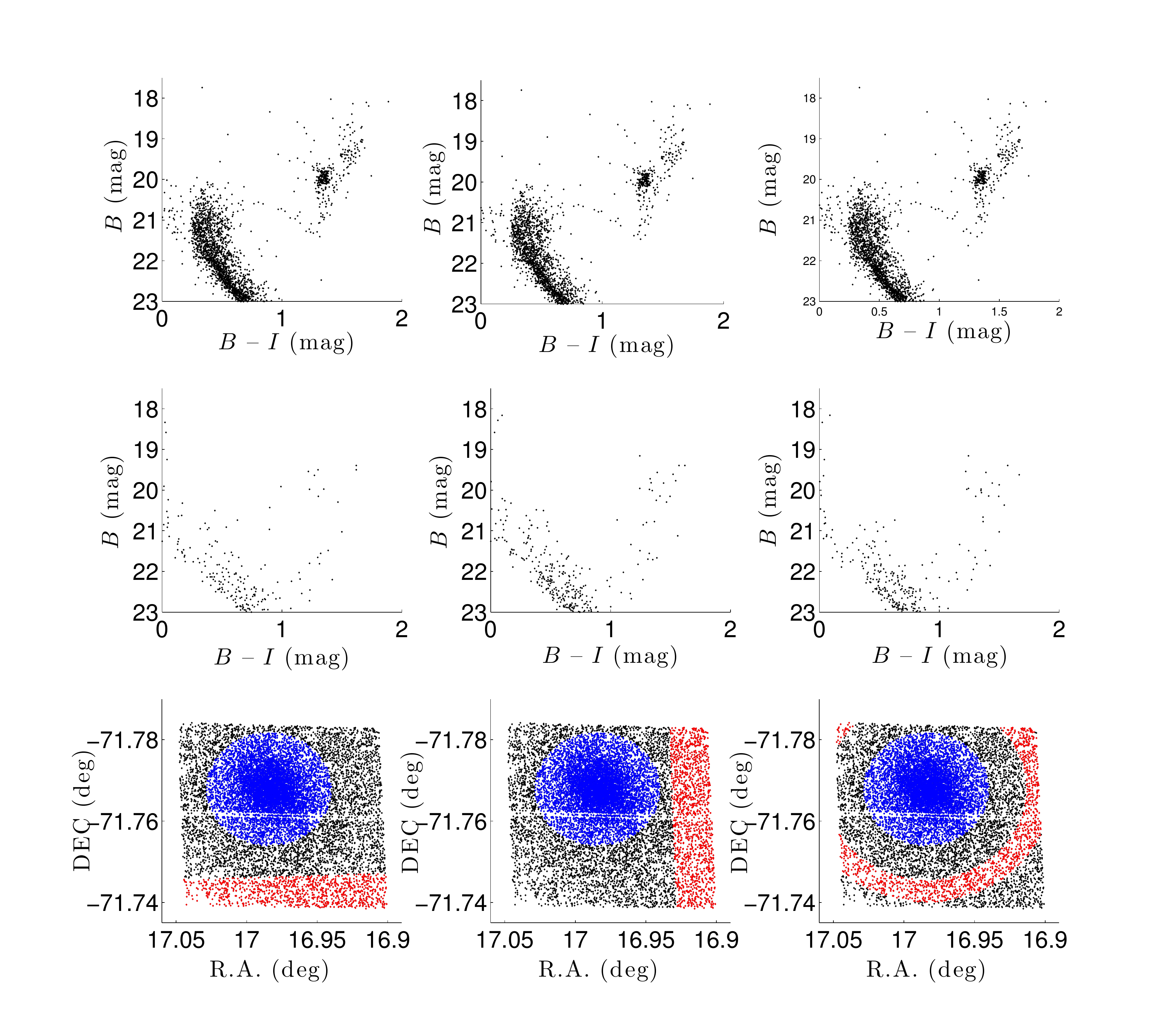}
\caption{Top row: Decontaminated CMDs based on different adopted
  background regions. The red rectangle covers the SGB region of
  interest. Middle row: Corresponding CMDs of the adopted field
  regions. Bottom row: Corresponding field regions (red dots) and
  cluster region (blue dots). Left: The background stars used have
  positions $Y$ $\le$ 800 pixels. Middle: The background stars have
  positions $X$ $\le$ 800 pixels. Right: The background stars are
  located between radii of 80 arcsec and 100 arcsec.}
\label{F1}
\end{figure*}

We have checked the effects of different adopted field
regions and found that the observed features, i.e., the tight SGB and
the eMSTO region, are independent of both the location and shape of
the field region. In Fig. \ref{F1} we present the decontaminated CMDs
(top panels) based on different adopted field regions (shown in the
bottom panels). For these different adopted field regions, the numbers
of field stars contaminating the SGB (RC) region of interest are two,
one and zero (five, ten and ten) for the fields shown in the bottom
left- to right-hand panels, respectively. We note that
  \cite{Cabr16a} have warned that a sparsely populated region may
  suffer from problems related to field-star contamination. Since the
  NGC 411 SGB region is such a sparsely populated region, we have
  tested if it could be populated mainly by field stars: we therefore
  adopted a similar approach to that of \cite[][their
    Fig. 4]{Cabr16a}. We thus confirmed that the average number
  dispersion of field stars in the SGB region does not exceed 1.34,
  while the number of stars detected in the SGB region is 18. Since
  this is at least 13 times the expected field-star level, this
  exercise thus supports our assertion that the observed SGB in NGC
  411 is a physical feature associated with the cluster rather than
  caused by field-star contamination. In addition, SMC field
  populations composed of stars with different ages and metallicities
  cannot form such a tight sequence in the CMD.

\section{Main Results}\label{S3}

\begin{figure*}
\centering
\includegraphics[width= 2\columnwidth]{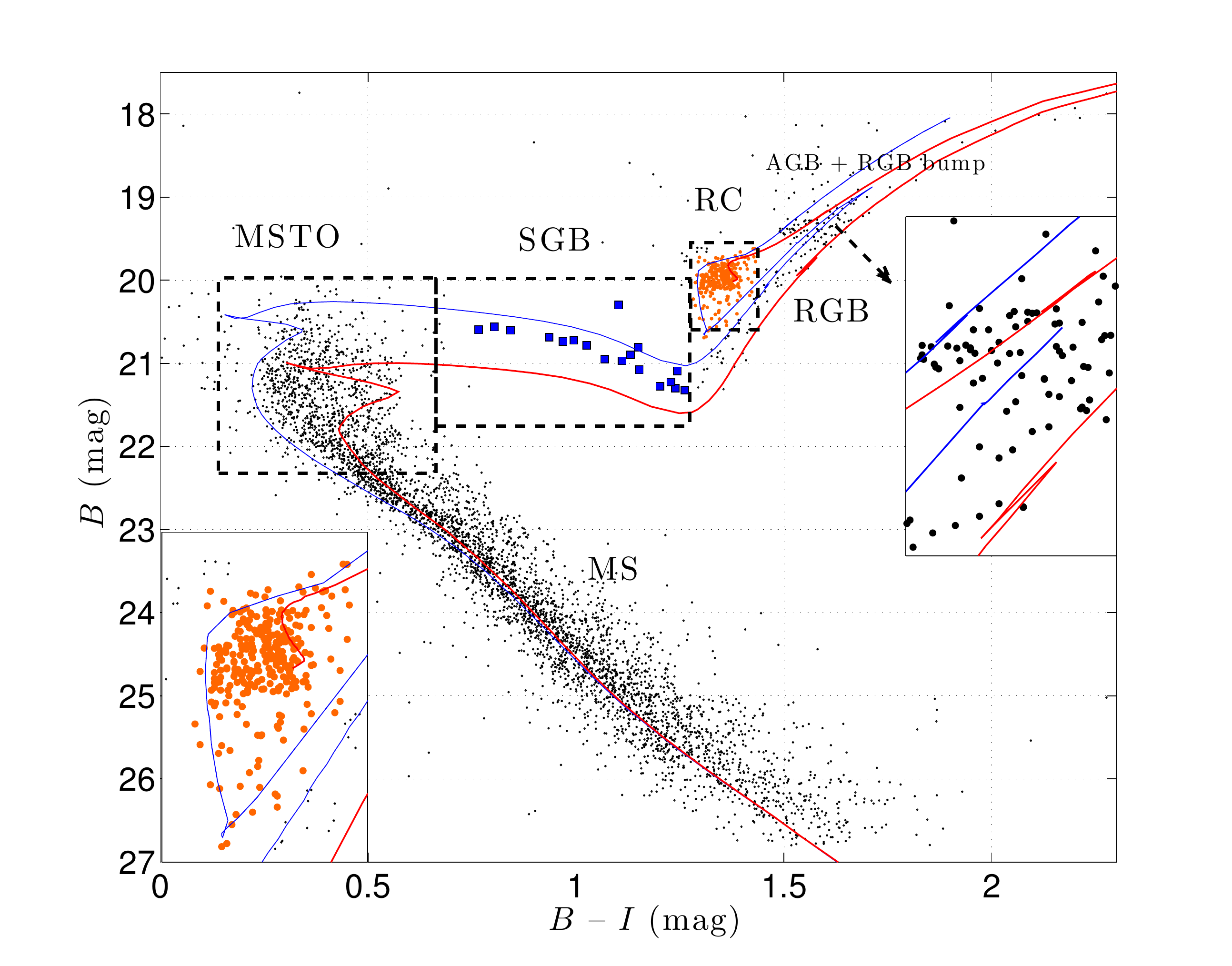}
\caption{NGC 411 CMD, showing isochrones for ages of 1.38 Gyr ($\log (t \mbox{
    yr}^{-1})$ = 9.14, blue solid line) and 2.18 Gyr ($\log (t \mbox{
    yr}^{-1})$ = 9.34, red solid line). The blue squares and orange dots
  represent the SGB stars and RC stars, respectively. A zoomed-in
  region focussing on the RC stars is included in the bottom left-hand
  panel. }
\label{F2}
\end{figure*}

Figure \ref{F2} shows the CMD of NGC 411 and its apparent eMSTO region
 for $20\le B \le 22$ mag. For stars brighter than $B =
21.5$ mag in the DAOPHOT \citep{Davi94} catalogue, the average
1$\sigma$ magnitude uncertainties are 0.04 mag in both bands. Clearly,
these small photometric errors cannot explain the broadening of the
eMSTO region. Below, we will discuss a CMD composed of a single
stellar population, assuming that only photometric uncertainties and
unresolved binary systems will affect the morphology of the MSTO
region. The cluster also exhibits both a well-populated SGB and a RC.
In Fig. \ref{F2} the SGB and RC stars analysed in this paper are
highlighted as blue rectangles and orange dots,
respectively \footnote{We also found a bump at $(B -
    I)\sim19.5$ mag and $B\sim1.6$ mag.  However, since a RC
  characterized by a significant age spread should not be dispersed so
  much in colour, it is more likely that this is a combination of a
  (weak) asymptotic-giant-branch (AGB) bump and a red-giant-branch
  (RGB) bump.}. Using the stellar evolutionary models of
\cite{Bres12}, we adopted a series of isochrones with ages of 1.38 Gyr
(${\log}$ ($t$ yr$^{-1}$) = 9.14) to 2.18 Gyr (${\log}$ ($t$
yr$^{-1}$) = 9.34) to describe the entire eMSTO region. Our fit
suggests a possible maximum age spread of $\sim$800 Myr, which is
slightly larger than that derived by \cite[][700
    Myr]{Gira13} but smaller than that found by \cite{Goud14}. All
other parameters pertaining to these isochrones are the same
as those adopted by \cite{Gira13}, including a
metallicity of $Z$ = 0.002, an extinction of $A_{V} = 0.25$ mag or
$E(B - V) = 0.08$ mag, and a distance modulus of $(m - M)_0 = 19.05$
mag. If an age spread were entirely responsible for the eMSTO, this
should be reflected in the width of the SGB. However, even at first
glance, we immediately see that the SGB is much narrower than would be
expected for an age spread of 800 Myr, we determined the best-fitting
age to the SGB is 1.58 Gyr (${\log}$ ($t$ yr$^{-1}$) = 9.20).

\begin{figure}
\centering
\includegraphics[width= 1\columnwidth]{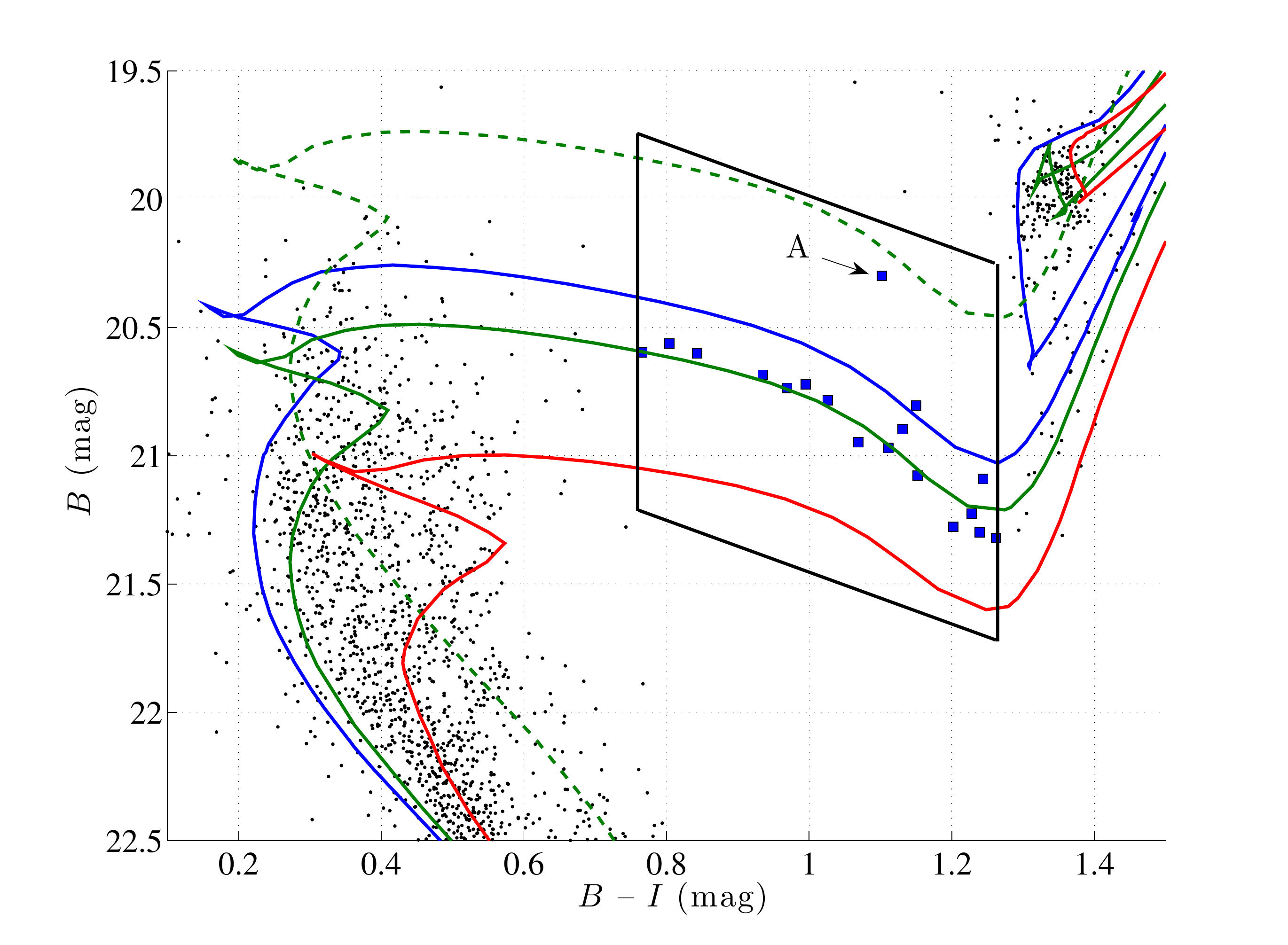}
\caption{Demonstration of our selection of SGB stars. The selected SGB
  stars (in the parallelogram-shaped box) are marked with blue
  squares, which are associated with the 1.58 Gyr isochrone (green solid line). 
  The green dashed line is the
  corresponding locus of equal-mass binary systems. Blue and
  red solid lines represent isochrones for 1.38 Gyr and 2.18 Gyr, respectively.}
\label{F3}
\end{figure}

Indeed, Fig. \ref{F2} already shows a very narrow SGB. However, since
this approach was disputed by \cite{Goud15}, who claimed that the
position of the stars along the SGB cannot properly constrain the
internal age distribution, and in order not to ignore any possible SGB
candidates, we compiled a sample of SGB stars by adopting a much
larger area of interest. We defined a parallelogram-shaped box,
approximately bounded by $19.7\le B \le 21.8$ mag and
  0.75 $\le$ ($B - I$) $\le$ 1.25 mag. This region was adopted to
  properly avoid possible contamination by stars located in the eMSTO
  region and at the bottom of the RGB. Stars located inside this box
were, hence, selected as possible SGB stars. This box spans more than
2 mag in brightness, a range that is sufficient to cover any
reasonable physical models one may consider. The maximum magnitude
offset caused by unresolved binaries is $\sim$0.75 mag and, as shown
by \cite{Goud15}, the enhancement in magnitude owing to stellar
convective overshooting (adopting overshooting ranging from
$\Lambda_{\rm e}$ = 0.35 to 0.50) is less than 0.1 mag. Figure
\ref{F3} shows the best-fitting isochrones to (i) the boundaries of
the eMSTO (blue dashed and red solid lines), (ii) the tight SGB (green
solid line) and (iii) the corresponding locus of equal-mass-ratio
unresolved binaries (green dashed line), it shows that only one star
(marked `A' in Fig. \ref{F3}) deviates significantly from the
best-fitting isochrone. As one can see, the luminosity of
  star `A' is much brighter than that of the youngest isochrone,
  characterized by an age of 1.38 Gyr or ${\log}$ ($t$ yr$^{-1}$) =
  9.14, which indicates that it maybe a bright foreground or background star. 
  It is also probably that star `A' is an approximately equal-mass unresolved 
  binary system, as its position is very close to the locus of the
  unresolved binary sequence. Both cases refuse that it is 
  an genuine single SGB star. Therefore, we have excluded this star from our
  analysis (in addition, a single exception will not affect our
  statistical results).



We next explore the morphology of the RC region to further constrain
the SFH. To select a statistically complete sample of RC stars, we
first select stars located in the colour and magnitude intervals
defined by $1.28\le(B - I)\le1.45$ mag and $19.6\le B \le
  20.7$ mag. Next, since the isochrones in the RC range trace open
loops, we adopt stars which are associated with the loop pertaining to
the most extreme isochrone used (normally, the 1.38 Gyr isochrone) as
RC stars. Of the stars located beyond this loop, we also select those
which are still located within the 2$\sigma$ uncertainties associated
with that isochrone.
Finally, we made a cut across the MSTO region to select the MSTO stars. 
This method is similar to the approaches used by \cite{Goud14}, \cite{Li14a} 
and \cite{Bast15}.

\begin{figure*}
\centering
\includegraphics[width= 2\columnwidth]{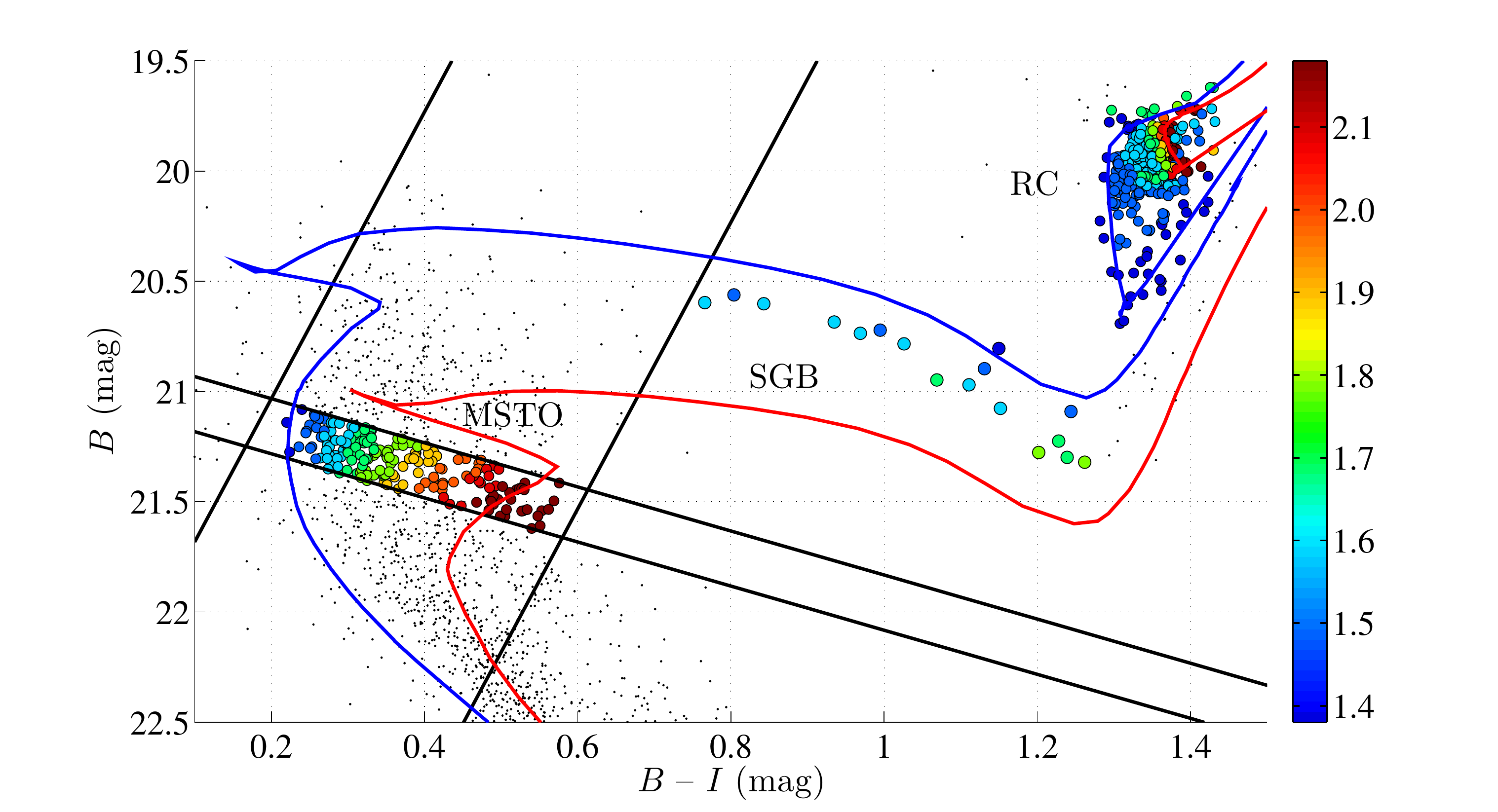}
\caption{CMDs of the NGC 411 MSTO, SGB and RC regions. The colours of
  the circles indicate their best-fitting ages (in Gyr).}
\label{F4}
\end{figure*}

To achieve the most straightforward comparison between the SFHs of the
eMSTO, SGB (without star `A') and RC, we directly assign the age of
the closest isochrone to each star \citep{Bast15}: see Fig. \ref{F4}.

\begin{figure}
\centering
\includegraphics[width = 1.0\columnwidth]{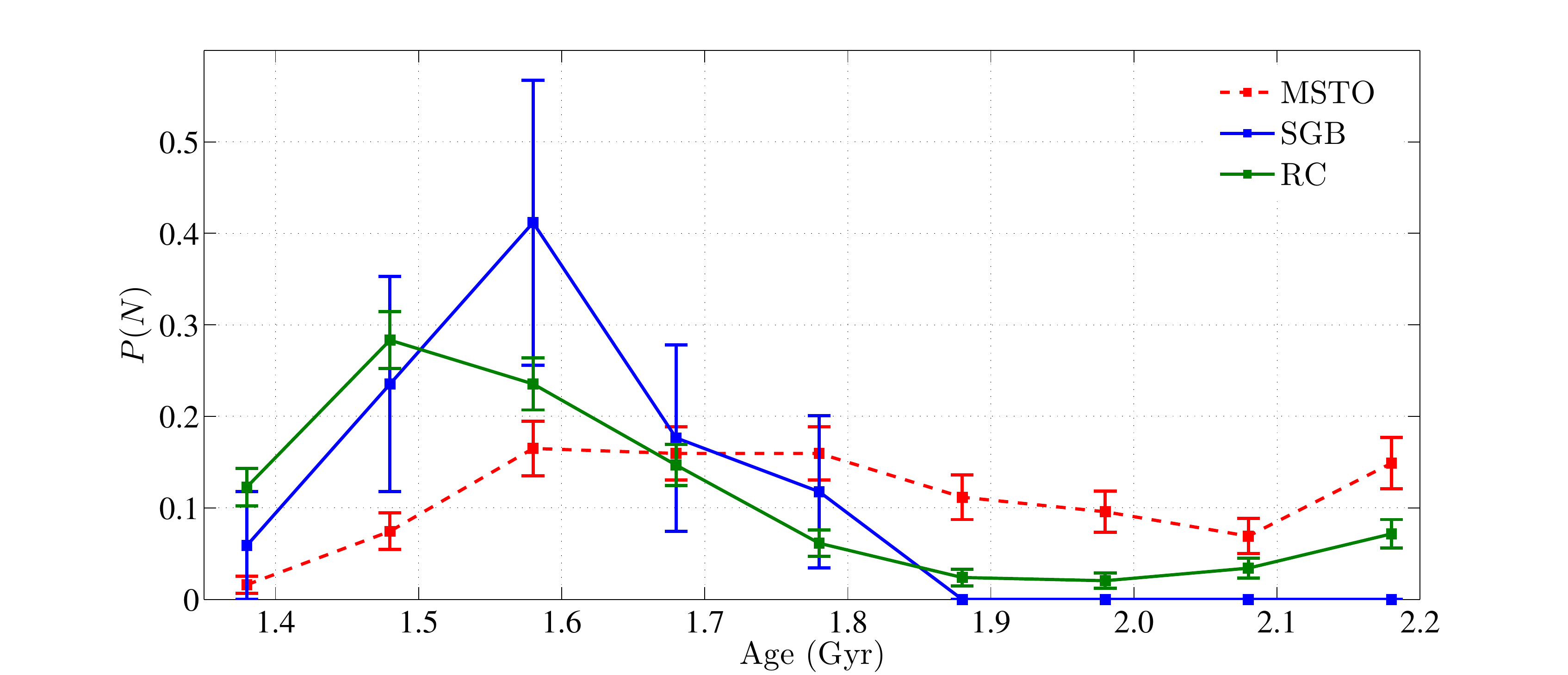}
\caption{SFHs derived from the MSTO stars (red), SGB stars (blue) and
  RC stars (dark green). All observed SGB stars and 85 percent of the
  cluster's RC stars appear to be younger than 1.78 Gyr. K--S-test
  results show that the SFHs of the SGB and RC stars are drawn from
  the same distribution, as opposed to the eMSTO-based SFH.
\label{F5}}
\end{figure}

In Fig. \ref{F5} we compare the SFHs derived from the SGB (without
star `A'), the RC and the eMSTO regions. The derived SFH for the eMSTO
region is almost flat from 1.38 Gyr to 2.18 Gyr, indicating an
apparent age spread. However, the SFHs resulting from both the SGB and
the RC stars preferentially render young ages: all SGB stars and 85
percent of the RC stars are younger than 1.78 Gyr. Using the K--S
test\footnote{The K--S test returns two values, $H$ and $P$. $H$ can
  only take the values 0 and 1, which depends on the null hypothesis
  that the two underlying distributions are independent. If $H$ is
  returned as 0, the null hypothesis is rejected, which means that the
  two samples under consideration are drawn from the same
  distribution. $P$ represents the $p$ value, which indicates the
  probability that the two samples are drawn from the same underlying
  population.} we quantify the similarities of the SFHs derived from
the SGB, the RC and the eMSTO. The K--S test reports that the SFHs
based on the SGB and the RC are all mutually consistent. However, when
we compare the SFHs derived from these features to that from the
eMSTO, the K--S test indicates that their SFHs are inconsistent with
that derived from eMSTO.
  
\begin{figure}
\centering
\includegraphics[width = 1.0\columnwidth]{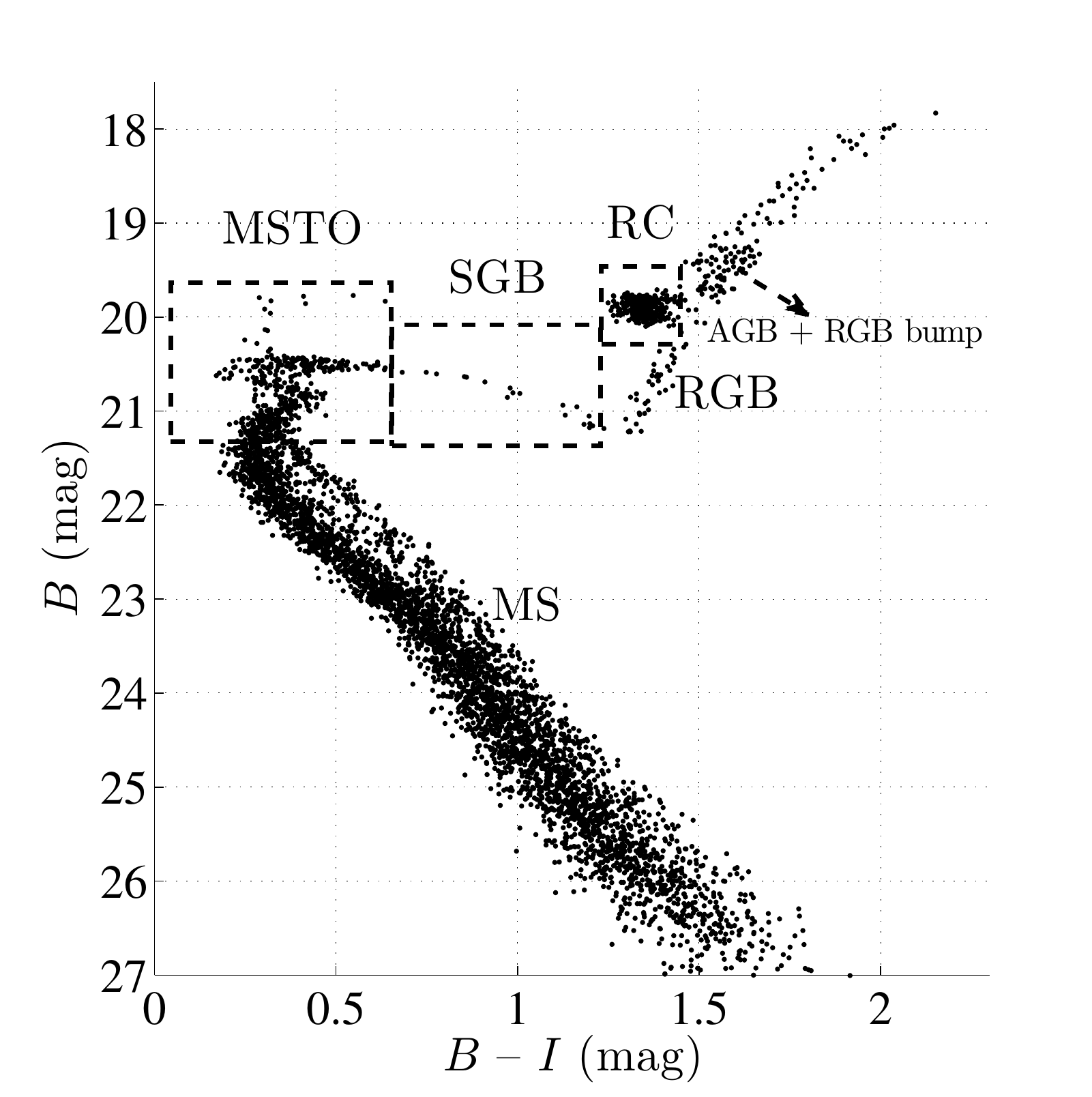}
\caption{Simulated CMD for an SSP with an age of ${\log}$ ($t$
  yr$^{-1}$) = 9.20.
\label{F6}}
\end{figure}

To test if their SFHs are consistent with SSPs, we generated a CMD of
an SSP with photometric uncertainties based on the 1.58 Gyr isochrone. 
We also evaluated the MS--MS binary
fraction for NGC 411 using a similar method to that used by
\cite{Milo12a}. We found that the fraction of binaries with mass
ratios $q \geq 0.6$ is approximately 20 percent. Assuming a flat
mass-ratio distribution \citep{Regg11}, we adopt a 50 percent MS--MS
binary fraction for our simulated CMD. For evolved stars we did not
assign binary status to any star, because only equal-mass binary
systems (i.e., SGB--SGB or RGB--RGB binaries) will exhibit significant
differences from single stars, and the numbers of such binary systems
are very small. The simulated MS stars have the same
luminosity function to the observations. We directly sample the evolved stars 
based on the Chabrier initial mass function \citep{Chab01a} which is built into the
isochrones\footnote{We assigned the SGB a similar colour distribution
  to the observations.}. Figure \ref{F6} shows the simulated CMD for 
  an SSP with an age of 1.58 Gyr.

\begin{figure*}
\centering
\includegraphics[width = 2.0\columnwidth]{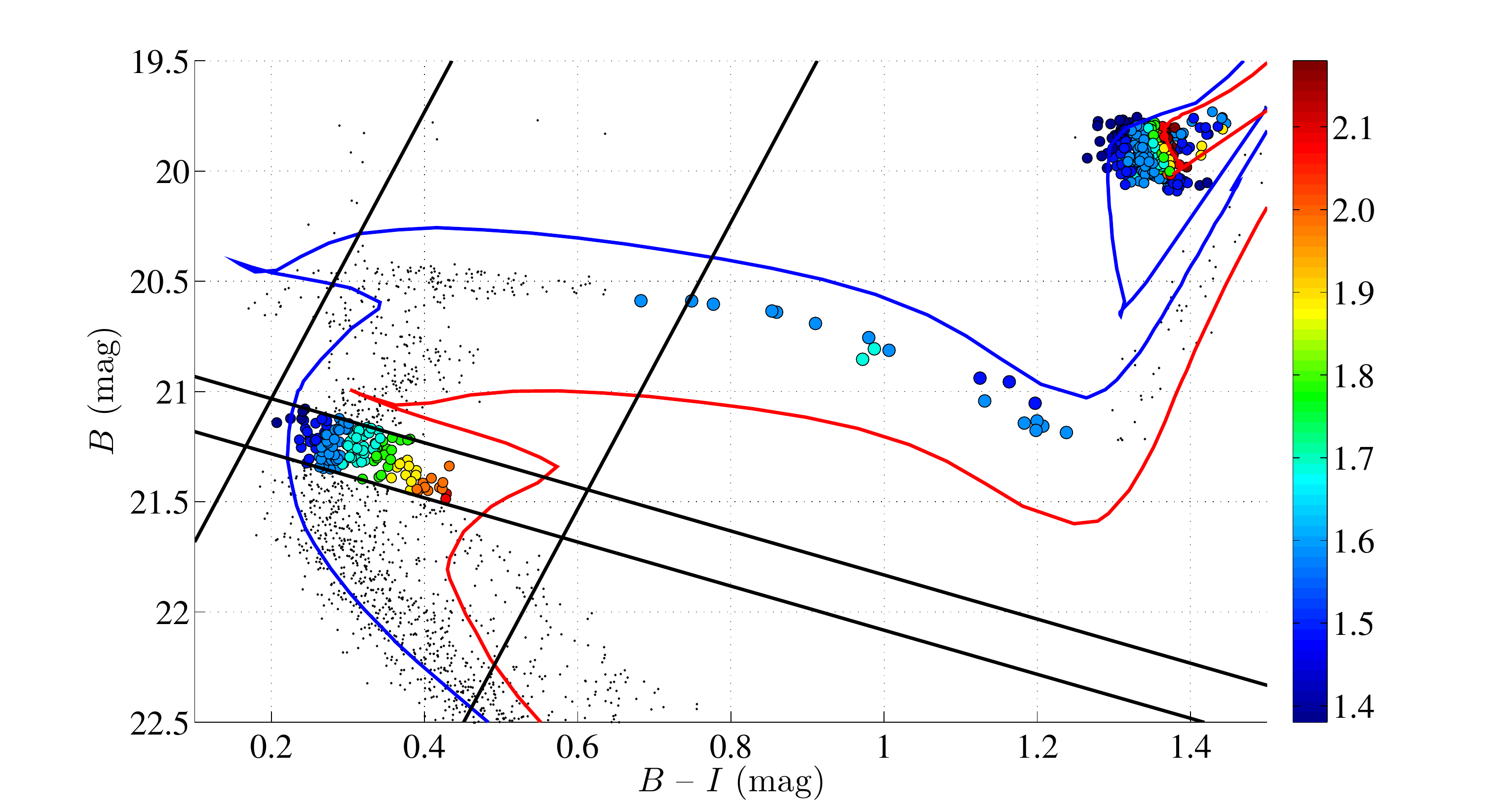}
\caption{Same as Fig. \ref{F4}, but for the simulated CMD of an SSP.
\label{F7}}
\end{figure*}

\begin{figure}
\centering
\includegraphics[width = 1.0\columnwidth]{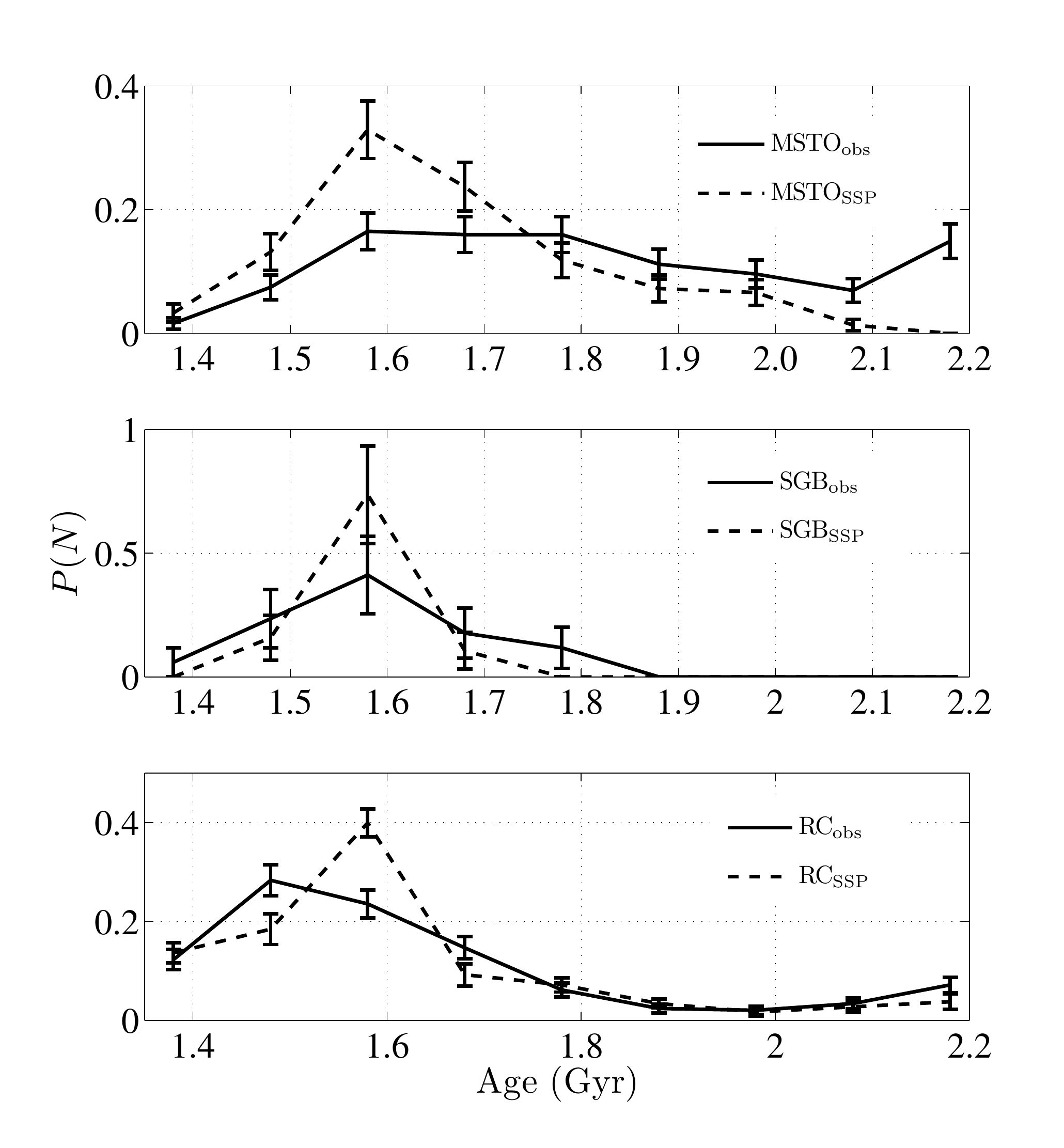}
\caption{Comparisons of the SFHs derived from the observations and the
  simulated SSP. Top panel: MSTO stars; middle: SGB stars; bottom: RC
  stars.
\label{F8}}
\end{figure}

Our next step applied to the simulated CMD is identical to the approach 
we performed on the observed CMD (see Fig.\ref{F7}). We derive the SFHs of the
simulated MSTO, SGB and RC stars and compare them with their
corresponding observations. The comparisons are shown in
Fig. \ref{F8}. Again, we use the K--S test to examine if the SFHs of
MSTO, SGB and RC stars are consistent with SSPs. The K--S-test results
show that the distributions of both the observed SGB and RC stars are
consistent with SFHs derived from SSPs ($P$ = 0.87 and 0.37,
respectively), but the observed MSTO is not consistent with an SSP:
($H,P$) = ($1, 6.8\times 10^{-8}$).
 
\begin{figure}
\centering
\includegraphics[width = 1.0\columnwidth]{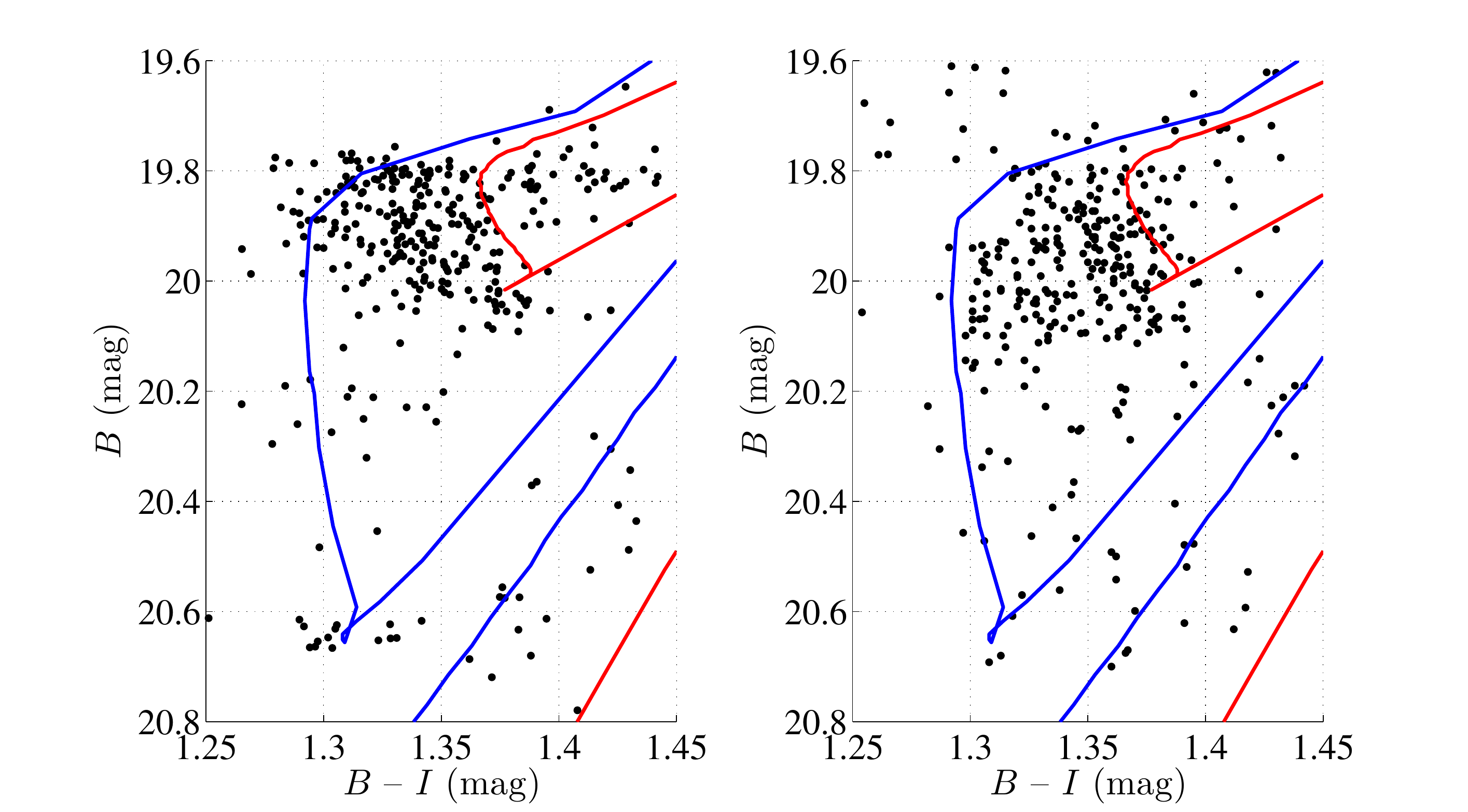}
\caption{Comparison of the synthetic RC (the synthetic stars include
  subpopulations characterized by ages of 1.38 Gyr and 1.48 Gyr, with
  number fractions of 12.5 percent each) and the observations. Left:
  synthetic RC region; right: observed RC region.
\label{F9}}
\end{figure}

However, based on visual inspection, we found a clear secondary RC
(SRC) which extends to $B \sim 20.2$ -- 20.5 mag (see
Fig. \ref{F4}). The more massive nature of the SRC stars may reflect a 
somewhat extended SFH, but it can be also be explained by rapid 
stellar rotation \citep{Taya13a,Dehe15a}. Clearly, in our simulation 
the RC of an SSP is much more compact than the observed RC 
(see Fig. \ref{F7}). We count the number of RC stars below the 
1.48 Gyr (${\log}$ ($t$ yr$^{-1}$) = 9.17) isochrone. We find that 
SRC stars represent $\sim$25 percent of the full RC population. 
To check whether the addition of younger populations to our models will 
improve the significance of the comparison, we added two populations 
with ages of 1.38 Gyr and 1.48 Gyr to 
the bulk population, with number fractions of 12.5 percent each.  Figure
\ref{F9} shows (left) the synthetic RC and (right) the observed
RC. The resulting morphologies of the MSTO and SGB are also slightly
different. We performed the same analysis as before, applied to the
synthetic MSTO, SGB and RC, and compared their derived SFHs with those
obtained for the observations: indeed, we found that once we add these 
two young populations, the K--S test again shows that the distributions 
of the observed SGB and RC stars are consistent with the SFHs used for the
synthetic samples, but with significant higher $P$ values of 0.93 and
0.67, respectively. The SFH of the eMSTO is still much more extended
than that of the synthetic sample, with ($H,P$) = ($1, 2.6\times
10^{-9}$). Based on this statistical analysis, we conclude that the SGB 
and RC morphologies do not agree with the eMSTO if age spreads are 
entirely responsible for the width of the eMSTO 
\footnote{Our adopted isochrones are the most up-to-date publicly
  available isochrones based on the PARSEC tracks \citep{Bres12}, with
  a mass resolution of $\sim$ 0.05 M$_{\odot}$, whereas the mass
  resolution of \cite{Goud15} is 0.01 M$_{\odot}$. However, their
  models are not publicly available.}.
  


The time-scales characteristic of the SGB phase for stars of different
masses deserve further discussion. From the best-fitting
isochrones, we found that the masses of SGB stars with an age of
1.38 Gyr are roughly 1.63 M$_{\odot}$, whereas
the masses of SGB stars with an age of 2.18 Gyr
are close to 1.38 M$_{\odot}$. Using the stellar evolutionary tracks
from \cite{Bert08a}, we estimated that the average time-scale for the
SGB phase of both of those types of stars are $\sim$15 Myr and 50 Myr,
respectively\footnote{The evolutionary tracks of \cite{Bert08a} only
  provide discrete physical parameters of stellar mass, helium
  abundance and metallicity. We selected $Y$ = 0.26 and $Z$ = 0.002
  for our estimates, which are the closest values to the physical
  parameters adopted here.} Clearly, if there is an age difference
among the observed SGB stars, then more faint SGB stars should be
expected. However, as shown in Fig. \ref{F3}, there are no SGB stars
associated with the old isochrone.


\section{Discussion}\label{S4}

\begin{figure}
\centering
\includegraphics[width = 1.0\columnwidth]{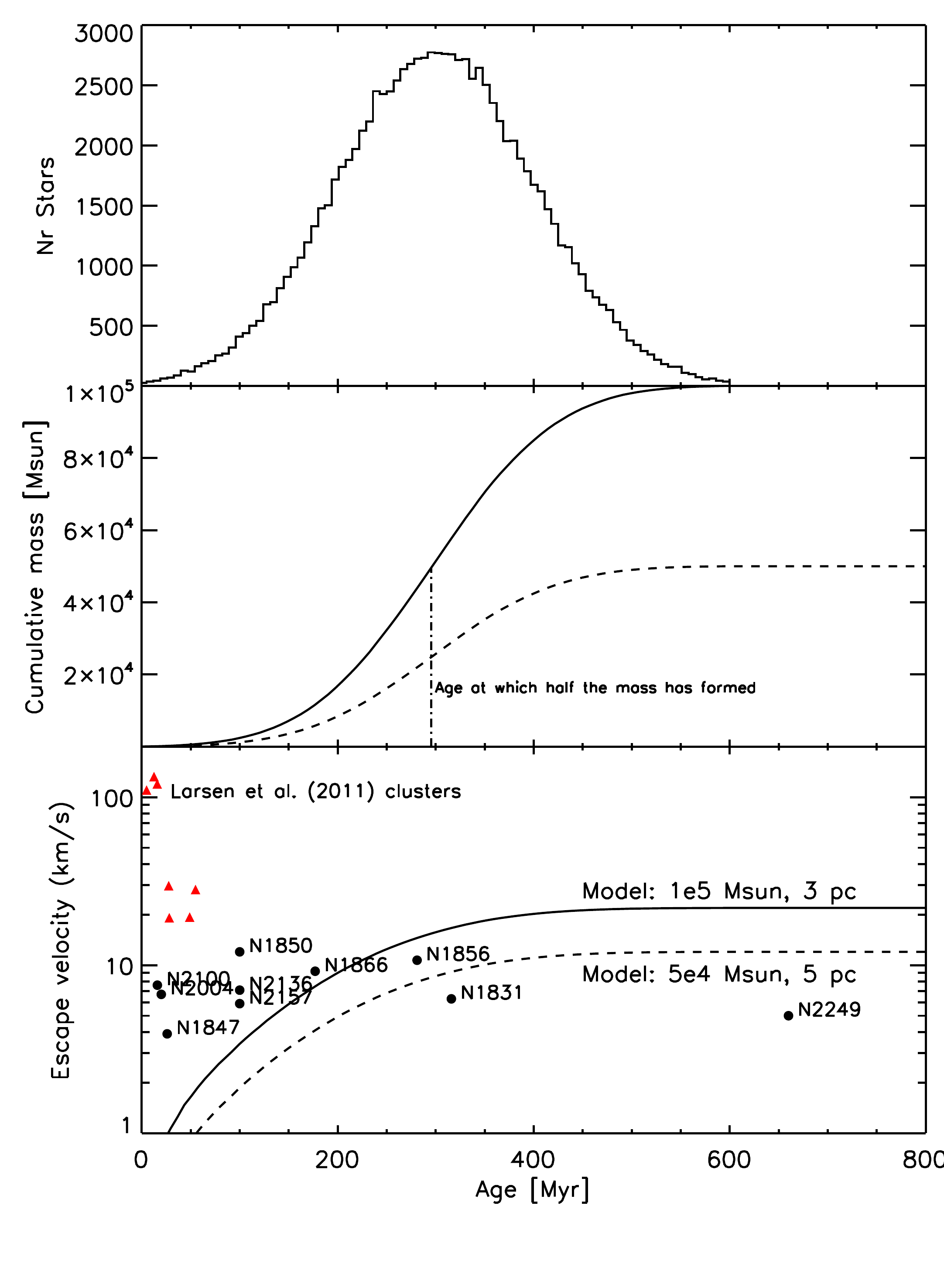}
\caption{Simulation of the build-up of stellar mass based on the
  derived SFHs of intermediate-age clusters. Top: Number of stars
  formed as a function of time for a $1\times10^5$ M$_{\odot}$ cluster
  with a Gaussian SFH characterized by $\sigma=100$~Myr and a peak at
  $300$~Myr. Middle: Corresponding cumulative build-up of mass for the
  same cluster (solid line) and for a cluster of $M_{\rm
    cl}=5\times10^4$ M$_{\odot}$ (dashed line). The age at which half
  the stellar mass is in place is shown by the vertical dash-dotted
  line. Bottom: Corresponding increase of the escape velocity of the
  clusters (where we have assumed a constant radius). Note that for
  such clusters it takes $\ge$ 100 Myr to reach an escape velocity
  greater than 10 km s$^{-1}$. Additionally, we show the loci of a
  sample of LMC and other young extragalactic clusters which do not
  exhibit ongoing star formation. These clusters do not host age
  spreads either which are consistent with those suggested for
  intermediate-age clusters based on their eMSTOs.
\label{F10}}
\end{figure}

\cite{Goud14} proposed a scenario which invokes two distinct
star-formation episodes in massive star clusters. In their model, the
second star-formation episode, which can last several hundred million
years following an initial near-instantaneous starburst, is
responsible for the observed eMSTO. Although the available
observations do not directly advocate a two-burst SFH for the clusters
thus far analysed -- i.e., only a single Gaussian SFH is inferred from
the eMSTO, in tension with the \cite{Goud14} scenario, while the
stellar age distribution in the young cluster NGC 1856 appears to be
inconsistent with such a scenario \citep[e.g.,][]{Milo15} -- this type
of scenario is required if significant age spreads are the cause of
the observed eMSTOs. This is so, because if the SFH directly inferred
from the eMSTO were adopted, this would imply that clusters initially
had very low masses (and thus a very low escape velocity) and needed
to build up their masses over the course of several hundred million
years. If this scenario were realistic, then we would expect to see
many young clusters with ages of 10 Myr to 100 Myr in the process of
building up their stellar masses. This is demonstrated in
Fig. \ref{F10}, where the required build-up of mass is compared with a
sample of LMC and other extragalactic clusters.

This scenario would result in an unphysical build-up of stellar mass,
i.e., it raises the question as to why a low-mass cluster with a low
velocity dispersion would be able to retain and accrete gas to form
second-generation stars. Additionally, the observed clusters do not
show evidence of ongoing star formation \citep{Bast13}, nor of the
build-up of their stellar mass over time in a way consistent with the
SFHs inferred from the eMSTOs. Hence, either the intermediate-age
eMSTO clusters are unique in their SFHs, or they did not build up
their stellar masses over hundreds of millions of years.

Another critical ingredient informing our discussion is that NGC 411
has a total mass of only ${\log}$ $M_{\rm cl}/{\rm M}_{\odot}$ $\sim$
4.51 \citep{McLa05}\footnote{Note that this value is lower than that
  adopted by \cite{Goud14}, because the latter authors use a single
  power-law stellar initial mass function instead of a more realistic
  Chabrier-type representation}. We can thus calculate $v_{\rm esc}$
\citep{Geor09},
\begin{equation}
v_{\rm esc} (t) = f_{\rm c}\,\sqrt{\frac{{M}_{\rm cl}(t)}{r_{\rm
      h}(t)}} \; \; \mbox{km s}^{-1},
\end{equation}
where $M_{\rm cl} (t)$ is the cluster mass at time $t$ in units of
M$_{\odot}$, $r_{\rm h} (t)$ is the cluster's half-light radius in pc
at time $t$ and $f_{\rm c}$ is a coefficient which depends on the
concentration index $c$ of \cite{King62} models. Based on the
assumptions that the half-mass radius of NGC 411 is roughly equal to
its half-light radius -- $r_{\rm h}$ = 6.1$\pm{0.8}$ pc \citep{Goud14}
-- and $f_{\rm c}$ $\sim$ 0.10 \citep[the average value derived from
  table 2 of][]{Geor09}, the resulting escape velocity for NGC 411 is
$v_{\rm esc}$ = 7.24$^{+2.92}_{-2.10}$ km s$^{-1}$, where the
uncertainty is based on a range of $f_{\rm c}$ from 0.075 to 0.130, as
well as on the uncertainty in $r_{\rm h}$. Clearly, the escape
velocity of NGC 411 is much lower than the threshold of 12--15 km
s$^{-1}$ proposed by \cite{Goud14}. These authors asserted that this
threshold is consistent with the wind velocities of intermediate-mass
AGB and massive binary stars.

In the \cite{Goud14} model, massive star clusters may have experienced
significant mass loss during their early evolutionary stages owing to
strong tidal stripping \citep{Vesp09}. If NGC 411 has lost a large
amount of its initial mass, this may suggest that its escape velocity
could have been much higher in the past. However, the current position
of NGC 411 is at a distance of approximately 11 kpc from the SMC's
centre \citep{Glat08}, and thus the strength of the tidal field it has
experienced is much lower than that in the simulation of \cite{DErc08}
adopted by \cite{Goud14}, in which clusters were located some 3 kpc
from the Milky Way centre. This is also discussed in detail in
\cite{Cabr15}

\subsection{Rapid stellar rotation}

\begin{figure*}
\centering
\includegraphics[width = 2.0\columnwidth]{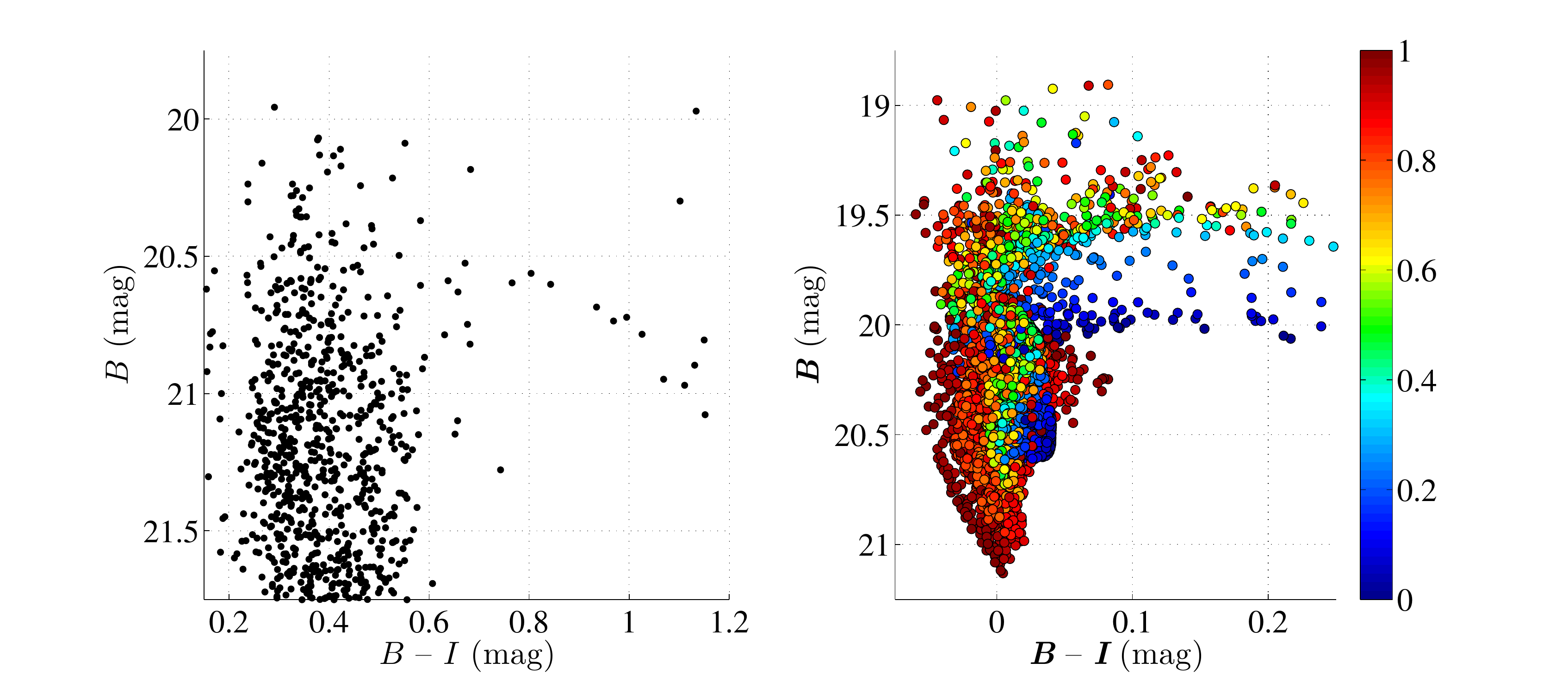}
\caption{Left: observed CMD of NGC 411; right: synthetic CMD composed
  of a sequence of fast rotators (with rotation rates indicated by the
  colour bar). Here the normal $B$ and $I$ labels represent the F475W
  and F814W magnitudes, whereas the bold-faced $B$ and $I$ labels
  represent magnitudes in the genuine Johnson--Cousins $B$ and $I$
  bands.
\label{F11}}
\end{figure*}

A number of researchers have shown that inferring age spreads in
intermediate-age star clusters on the basis of their eMSTO morphology
is likely invalid if many MSTO stars are fast rotators
\citep{Bast09,Yang13,Li14a,Li14,Bran15}. The overall effect of 
rapid stellar rotation is complicated. Many small contributions (e.g., gravity 
darkening, stellar rotational mixing, hydrodynamic effects, etc.) may 
result in a not-so-simple morphology of their colour--magnitude distribution. 
A reasonable expectation is that a combination of those effects 
should be at work in NGC 411. If rotational mixing also plays a 
role in broadening the MSTO region, this would still give rise to a 
broadened or split SGB \citep[for a discussion, see][]{Li14}. To obtain 
a comparison of our observations with numerical calculations, 
we used the Geneva population synthesis code \citep{Ekst12,Geor14} to 
generate a CMD including stars with a range of rotation rates, 
based on their interactive tools.\footnote{http://obswww.unige.ch/Recherche/evol/-Database-}
Their models are available for nine different rotation rates
($\Omega_{\rm ini}$/$\Omega_{\rm crit}$ = 0.0, 0.1, 0.3, 0.5, 0.6,
0.7, 0.8, 0.9 and 0.95, where $\Omega$ is the stellar surface angular
velocity and $\Omega_{\rm crit}$ is the stellar break-up angular
velocity \cite{Geor14}), but their models only work for stars with
masses above 1.7 M$_{\odot}$. Figure \ref{F11} shows a comparison of
the observations (left) and the synthetic CMD (right). The latter
contains stars with various initial rotation rates and most of the
adopted physical parameters are equal to our previous fits: we adopted
a metallicity of $Z$ = 0.002 and a distance modulus of $(m - M)_0 =
19.05$ mag. However, note that the default extinction in the Geneva
models is fixed at $E(B-V)$ = 0 mag. The photometric magnitude system
of the Geneva models is the genuine Johnson--Cousins system, which is
different from that of our observations (which uses the {\sl
  HST}/WFC3 UVIS magnitude system). As shown in the right-hand panel of
Fig. \ref{F11}, SGB stars with different initial rotation rates
(indicated by the colour bar) produce a clear, extended SGB. Clearly,
this is not found in our observations.

The inconsistency between the simulated CMD and the observations thus
leaves us with a conundrum, since it seems that our currently favoured
solution is invalid. However, the age of NGC 411 reaches 1.58 Gyr,
while currently the Geneva models can only extend down to age of
${\log}$ ($t$ yr$^{-1}$) = 9.00 (1 Gyr) for the physical parameters
adopted here. This means that we cannot make a direct
comparison. Additionally, we note that in the Geneva models, stars
that rotate (even with small values of $\Omega$ $\sim$ 0.2 and higher)
converge to occupy a narrow SGB. It is only non-rotating stars that
would be expected to form a second sequence that have luminosities
below the main SGB. The effects of rapid stellar rotation are very
complicated, since rotational mixing may also lead to chemical
anomalies. The discovery of nitrogen and helium enhancements in
massive MS stars has generally been recognised as being caused by
rotational mixing \citep{Scho88,Gies92}. Enhanced helium abundances,
as well as increased metallicities, will impact the morphology of the
observed CMD \citep[e.g.,][]{Piot07a,Milo12b}; see also
  the RC discussion of \cite{Cole98a}. It is not entirely clear what
the rotational scenario would predict for the RC. For younger
clusters, \cite{Nied15} showed that the RC is expected to be narrower
than in the case of real age spreads, but still more extended than an
SSP, this is also consistent with our observations of SRC. The RC is
also metallicity-dependent, and rapid stellar rotation may also affect
the patterns of the stellar surface chemical abundance. To understand
the eMSTO conundrum, more detailed studies of the effects of rapid
stellar rotation, as well as of binarity fractions in intermediate-age
star clusters are required. Altogether, at this stage, we cannot
directly compare the models and observations before we extend the
stellar evolutionary tracks for different rotational rates to lower
masses.

\section{Conclusions}\label{S5}

In this paper, we have tested the interpretation of an internal age
spread as the possible explanation of the eMSTO morphology in NGC 411
by exploring its SGB. Our results show that the width of the SGB in
NGC 411 does not suggest any significant age spread, despite the fact
that it exhibits an apparent eMSTO region with a putative spread of up
to 1 Gyr. Except for a single star, we found that the observed SGB is
well fitted by a simple-aged stellar population of 1.58 Gyr, which is 
not consistent with the age distribution inferred from the eMSTO 
region. We also found that the SFH derived from the RC is 
consistent with that obtained from the SGB; most RC and
SGB stars are younger than 1.78 Gyr. Our comparison among the SFHs
derived using SGB, RC and eMSTO stars shows that they do not support
an age spread of $\sim$ 800 Myr as derived from the eMSTO region.
  
We conclude that the observed tight SGB and RC exclude the presence of
a significant age spread in NGC 411. The scenario proposed by
\cite{Goud14}, which invokes two distinct star-formation events to
build up the stellar mass of star clusters within several hundred
million years, seems unphysical for NGC 411, since its escape velocity
is much lower than that suggested by \cite{Goud14} and NGC 411 has
unlikely undergone significant mass loss in the past. The presence of
an eMSTO in NGC 411 cannot be explained by an extended SFH. This is
also consistent with the recent study of \cite{Piat16} who found
eMSTOs in two intermediate-age LMC clusters with masses of $\sim$ 5000
M$_{\odot}$.

However, we cannot directly compare the simulations with our
observations of NGC 411 based on the stellar rotation scenario because
the models do not currently extend to this mass range. A direct
explanation of the observed eMSTO, combined with the tight SGB, is
that the eMSTO may be caused by rapid stellar rotation, whereas fast
rotators would quickly slow down when they evolve off the MS owing to
the conservation of angular momentum. However, rotational mixing may
also complicate the morphology of the SGB. We assert that to
understand the observed eMSTOs in intermediate-age star clusters, more
detailed investigations of rapid stellar rotation, as well as its
effects on stellar surface abundances, are required.

\section*{Acknowledgements}
We thank Sylvia Ekstr\"om for assistance and discussions. 
C. L. is partially supported by the Strategic Priority Programme `The Emergence
of Cosmological Structures' of the Chinese Academy of Sciences (grant
XDB09000000) and by the Macquarie Research Fellowship Scheme.
R. d. G. and L. D. acknowledge funding from the National Natural
Science Foundation of China (NSFC; grants 11073001, 11373010 and
11473037). N. B. is partially funded by a Royal Society University
Research Fellowship and an European Research Council Consolidator
Grant (Multi-Pop; 646928). F. N. was supported by the DFG Cluster of
Excellence `Origin and Structure of the Universe'.  C.Z. and his advisor 
Chen Li would acknowledge the support by ``973 Program'' 2014 CB845702, 
the Strategic Priority Research Program ``The Emergence of Cosmological 
Structures'' of the Chinese Academy of Sciences (CAS; grant XDB09010100)

\bibliographystyle{mnras}
\bibliography{cz16c}

\bsp	
\label{lastpage}
\end{document}